\documentstyle[12pt,epsf]{article}
\def\appendix#1{
  \addtocounter{section}{1}
  \setcounter{equation}{0}
  \renewcommand{\thesection}{\Alph{section}}
 \section*{Appendix \thesection\protect\indent \parbox[t]{11.715cm} {#1}}
  \addcontentsline{toc}{section}{Appendix \thesection\ \ \ #1}
  }

\newcommand{\newsection}{
\setcounter{equation}{0}
\section}

\def\bea{\begin{eqnarray}}
\def\eea{\end{eqnarray}}
\def\be{\begin{equation}}
\def\ee{\end{equation}}
\newcommand{\tr}[1]{\:{\rm tr}\,#1}
\newcommand{\Tr}[1]{\:{\rm Tr}\,#1}

\def\e{{\,\rm e}\,}
\def\d{\partial}
\def\D{\delta}

\newcommand{\rf}[1]{(\ref{#1})}
\newcommand{\non}{\nonumber \\*}
\def\g{g^2_{\rm YM}}
\def\fc{\Phi_{\rm cl}}
\def\diag{{\rm diag}}
\def\a{\alpha}

\begin{document}
\begin{titlepage}
\begin{flushright}
ITEP--TH--3/99
\end{flushright}
\vspace{1.5cm}

\begin{center}
{\LARGE Coleman-Weinberg Mechanism} 
\\[.5cm]
{\LARGE and Interaction of D3-Branes in Type 0 String Theory}\\
\vspace{1.9cm}
{\large K.~Zarembo}\\
\vspace{24pt}
{\it Department of Physics and Astronomy,}
\\{\it University of British Columbia,}
\\ {\it 6224 Agricultural Road, Vancouver, B.C. Canada V6T 1Z1} 
\\ \vskip .2 cm
and\\ \vskip .2cm
{\it Institute of Theoretical and Experimental Physics,}
\\ {\it B. Cheremushkinskaya 25, 117259 Moscow, Russia} \\ \vskip .5 cm
E-mail: {\tt zarembo@theory.physics.ubc.ca/@itep.ru}
\end{center}
\vskip 2 cm
\begin{abstract}
The low-energy theory on the world volume of 
parallel static D3-branes of type 0 strings 
is the Yang-Mills theory with six scalar fields in the adjoint 
representation. One-loop corrections in this theory
induce Coleman-Weinberg effective
potential, which can be interpreted as an interaction energy
of D3-branes.
The potential is repulsive at short distances and attractive at large
ones. In the equilibrium, a large number of D3-branes forms
a spherical shell with the radius proportional to the characteristic
energy scale of the world-volume theory.
\end{abstract}

\end{titlepage}

\setcounter{page}{2}

\newsection{Introduction}

The duality of ${\cal N}=4$, $D=4$ supersymmetric Yang-Mills theory
to the type IIB string theory on the 
near-horizon geometry of the three-brane
\cite{mal97,gkp98,wit98} have led to considerable progress in
understanding of its large-$N$ limit. Along the same lines, type 0B
theory on the background of RR charged three-brane solution was 
proposed to give a dual description of non-supersymmetric $D=4$ gauge 
theory coupled to adjoint bosonic matter \cite{kt98}. Perturbatively
unstable tachyon of type 0 theory was argued to be stabilized in the
presence of the three-brane through interaction with background
RR flux.  

The low-energy theory on the world volume of $N$
coincident type 0B D3-branes \cite{bg97} is $U(N)$ gauge theory 
with six scalar fields in the adjoint representation. According to 
\cite{kt98}, this theory has a dual description in terms of type 0B 
strings on the background of the three-brane. The classical gravity 
approximation to the dual picture already involves qualitative
features expected from the gauge theory, such as logarithmic dependence
of the coupling on a scale with UV fixed point at zero coupling
\cite{min98,kt98'}. This duality 
also predicts IR fixed point at infinity \cite{kt98'}.

The gauge theory considered in \cite{kt98,min98,kt98'} has the same
tree-level bosonic action as ${\cal N}=4$ SYM theory. So, the scalar
potentials in both theories have the same flat directions. These flat
directions correspond to transverse coordinates of D3-branes. But, unlike
in ${\cal N}=4$ theory, in the  
non-supersymmetric case the flat directions are not protected
from being lifted by quantum corrections, which reflects the fact that
parallel D3-branes of type 0 strings interact with one another, while
type II D-branes are BPS states and they can be moved apart at no
energy cost. Qualitative arguments based on the string
calculation of the interaction potential suggest that type 0 branes 
attract at large distances \cite{kt98}.

We study the interaction between type 0 D3-branes  
computing the one-loop effective potential in the world-volume field
theory. Similar calculations for ${\cal N}=4$ SYM theory with
the supersymmetry broken by the finite temperature 
\cite{wit98'} were done in \cite{ty98}. We find that the potential
has a maximum at zero separation between branes (at zero expectation 
values of scalar fields) and gains a minimum at finite separation
due to Coleman-Weinberg mechanism \cite{cw73}.

\newsection{Interaction potential}

The tree-level action of the low-energy theory on the 
world volume of $N$ 
parallel D3-branes of the type 0 string theory is
\be\label{act}
S=\frac{1}{\g}\int d^4x\,\tr\left\{-\frac12\,F_{\mu\nu}^2
+\left(D_\mu\Phi^i\right)^2-\frac12\,[\Phi^i,\Phi^j]^2\right\}.
\ee
We consider the theory in the Euclidean space from the very beginning. 
Strictly speaking, the field theory with the action
\rf{act} is not renormalizable and require counterterms quadratic and 
quartic in scalar fields. In what follows we imply that all
necessary counterterms are added to the action.

The scalar potential in \rf{act} has a degenerate set of minima:
\be\label{cl}
\fc^i=\diag(y_a^i),~~~a=1,\ldots, N.
\ee
The coordinates $y^i_a$, $i=1,\ldots, 6$ describe positions of
$N$ parallel static three-branes in nine-dimensional space.
Since the potential in \rf{act} does not depend on $y_a^i$, D-branes 
do not interact at the classical level. 

However, one-loop corrections induce interaction between branes
via the Coleman-Weinberg mechanism.
We calculate the interaction potential expanding 
scalar fields around the classical background \rf{cl}:
\be
\Phi^i=\fc^i+\phi^i
\ee
and integrating out quantum fluctuations. To integrate over the
gauge fields, we add to the action the gauge fixing term:
\be
S_{\rm gf}=-\frac{1}{\g\a}\int d^4x\,\tr\left(\d_\mu A_\mu
-\a[\fc^i,\Phi^i]\right)^2,
\ee
where $\a$ is a gauge fixing
parameter. The action for ghosts in the chosen gauge 
is
\be
S_{\rm gh}=\frac{1}{\g}\int d^4x\,\tr\left(\d_\mu\bar{c}D_\mu c
-\alpha[\fc^i,\bar{c}][\Phi^i,c]\right).
\ee
Expanding the action to the second order in fluctuations 
and integrating them out we get the
one-loop effective potential:
\bea
\Gamma&=&\frac12\,\Tr\ln\left[\left(-\d^2+Y^2\right)\D_{\mu\nu}
+\left(1-\frac{1}{\a}\right)\d_\mu\d_\nu\right]
-\Tr\ln\left(-\d^2+\a Y^2\right)
\non &&
+\frac12\,\Tr\ln\left[\left(-\d^2+Y^2\right)\D^{ij}-(1-\a)Y^iY^j\right].
\eea
The first term is the contribution of the gauge fields, 
the second is that of the ghosts, and the third of the scalars. 
By $Y^i$ we denote the following
matrix in the adjoint representation of $U(N)$:
\be
Y^i=[\fc^i,\cdot].
\ee
Taking into account that $[Y^i,Y^j]=0$, we find:
\bea
\Gamma&=&4\Tr\ln\left(-\d^2+Y^2\right)={\rm Vol}\,\,
4\int\frac{d^4p}{(2\pi)^4}
\,\tr\ln\left(p^2+Y^2\right)
\non 
&=&\mbox{quadratically divergent term}+
{\rm Vol}\,\,\frac{1}{8\pi^2}\,\tr Y^4\ln
\frac{Y^2}{M^2},
\eea
where $M$ is an UV cutoff. The quadratic and the logarithmic 
divergencies in the effective action should be canceled by
appropriate counterterms. 

The matrix $Y^2$ has eigenvalues $(y_a-y_b)^2$, so the one-loop corrections
induce only two-body interactions of D-branes -- the interaction 
potential is
\be
\Gamma={\rm Vol}\,\,\frac{1}{4\pi^2}\sum\limits_{a<b}|y_a-y_b|^4\ln
\frac{|y_a-y_b|^2}{\Lambda^2},
\ee
where $\Lambda$ is a non-perturbative mass scale of the world-volume theory.

\newsection{Equilibrium configuration of D-branes.}

\begin{figure}[t]
\hspace*{5cm}
\epsfxsize=7cm
\epsfbox{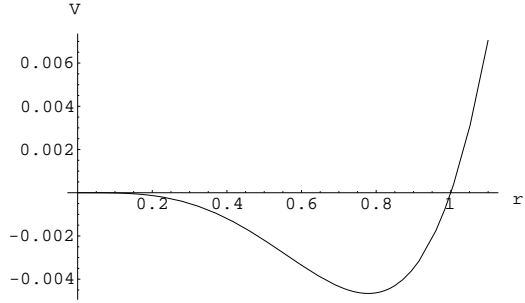}
\caption[x]{Interaction potential \protect\rf{int} in the units of $\Lambda$.}
\label{figpot}
\end{figure}
The potential of interaction between two D-branes\footnote{We use the
string units, $\alpha'=1$.}
 (fig.~\ref{figpot}):
\be\label{int}
V(r)=\frac{1}{4\pi^2}\,r^4\ln\frac{r^2}{\Lambda^2},
\ee
is such that D-branes repulse at short distances. Thus, a stack of 
D-branes put 
on top of each other is unstable. The D-branes will tend to separate
by distances of order $\Lambda$. 
In the equilibrium, transverse coordinates of
D-branes satisfy the equations:
\be\label{eq}
\sum\limits_b(y_a^i-y_b^i)|y_a-y_b|^2
\ln\frac{|y_a-y_b|^2}{\Lambda^2\e^{-1/2}}=0.
\ee
We are interested in the case when the number of D-branes, $N$, is large.
In the $N\rightarrow\infty$ limit, D-branes form continuous
spherically symmetric distribution and the equation \rf{eq}
takes the form:
\be\label{eq'}
y_a^if(|y_a|^2)=0.
\ee
The function $f$, in principle, can have several zeros, but we adopt
rather natural assumption
that for the configuration of D-branes which has a minimal energy the 
equation \rf{eq'} has only one root, $|y|^2=R^2$. So, the D-branes
in the equilibrium will form a spherical shell of radius $R$ in the 
six-dimensional transverse space with a surface RR charge density 
$\rho=N/\pi^3R^5$. From eqs.~\rf{eq},~\rf{eq'} we find:
\be
R=\Lambda\e^{-\frac{589}{840}},
\ee 
and the interaction energy per unit volume, 
which shifts the tension of a D-brane, is
\be
\Delta T=-\frac{7N^2\Lambda^4}{48\pi^2}\,\e^{-\frac{589}{210}}.
\ee

\newsection{Discussion}

We have calculated the interaction potential between D3-branes of
type 0B string theory at weak coupling in the world-volume
field theory\footnote{The definition of the YM coupling in terms
of VEVs of the dilaton and the tachyon is discussed in \cite{kt98',gar99}.}.
As long as the one-loop approximation can be trusted, the
field theory predicts the repulsion of type 0 D3-branes at short
distances and the attraction at large ones. As a result, D3-branes
tend to spread over the distances that are determined by a characteristic
scale of the low-energy world-volume theory. Such behavior is not
expected in the case of dyonic branes discussed in \cite{kt99}, because
the field theory on their world volume is conformal.

\subsection*{Acknowledgments}

This work was supported by NATO Science Fellowship and, in part, by
 INTAS grant 96-0524,
 RFFI grant 97-02-17927
 and grant 96-15-96455 of the promotion of scientific schools.


\end{document}